\documentclass[reqno]{amsart}

\usepackage{booktabs} 
\usepackage{IEEEtrantools}
\usepackage{amssymb,latexsym,amsfonts,amsmath}
\usepackage{graphicx}
\usepackage{dsfont}

\usepackage{algorithm}
\usepackage{algorithmic}

\usepackage{tikz}
\usetikzlibrary{calc,shapes,arrows}

\newtheorem{theorem}{Theorem}[section]
\newtheorem{lemma}[theorem]{Lemma}

\newtheorem{proposition}[theorem]{Proposition}

\newtheorem{definition}[theorem]{Definition}

\newtheorem{remark}[theorem]{Remark}
\newtheorem{assumption}[theorem]{Assumption}
\numberwithin{equation}{section}

\newcommand{\R}{{\mathbb{R}}}

\newcommand{\N}{{\mathbb{N}}}

\begin{document}

\begin{abstract}
In this paper, we propose a compositional framework for the construction of control barrier certificates for large-scale stochastic switched systems accepting \emph{multiple} control barrier certificates with some \emph{dwell-time} conditions. The proposed scheme is based on a notion of so-called \emph{augmented pseudo-barrier certificates} computed for each switched subsystem, using which one can compositionally synthesize state-feedback controllers for interconnected systems  enforcing safety specifications over a finite-time horizon. In particular, we first leverage sufficient $\max$-type small-gain conditions to compositionally construct augmented control barrier certificates for interconnected systems based on the corresponding augmented pseudo-barrier certificates of subsystems. Then we quantify upper bounds on exit probabilities - the probability that an interconnected system reaches certain \emph{unsafe} regions - in a finite-time horizon using the constructed augmented barrier certificates. We employ a technique based on a counter-example guided inductive synthesis (CEGIS) approach to search for  control barrier certificates of each mode while synthesizing safety controllers providing switching signals. We demonstrate our proposed results by applying them first to a room temperature network containing $1000$ rooms. Finally, we apply our techniques to a network of $500$ switched subsystems (totally $1000$ dimensions) accepting \emph{multiple} barrier certificates with a \emph{dwell-time} condition, and provide upper bounds on the probability that the interconnected system reaches some unsafe region in a finite-time horizon. 
\end{abstract}

\title[Compositional Control Barrier Certificates for Stochastic Switched Systems]{Compositional Construction of Control Barrier Certificates for Large-Scale Stochastic Switched Systems}

\author{Ameneh Nejati$^1$}
\author{Sadegh Soudjani$^2$}
\author{Majid Zamani$^{3,4}$}
\address{$^1$Department of Electrical and Computer Engineering, Technical University of Munich, Germany.}
\email{amy.nejati@tum.de}
\address{$^2$School of Computing, Newcastle University, United Kingdom}
\email{sadegh.soudjani@newcastle.ac.uk}
\address{$^3$Department of Computer Science, University of Colorado Boulder, USA}
\address{$^4$Department of Computer Science, Ludwig Maximilian University of Munich, Germany}
\email{majid.zamani@colorado.edu}
\maketitle

\section{Introduction}

{\bf Motivations.}
This paper is mainly motivated by the challenges emerging in the control of large-scale stochastic switched systems. In the past few years, stochastic switched systems have obtained considerable attentions among both control and computer scientists due to their broad applications in modeling many real-life systems. Since the closed-form solution of synthesized policies for stochastic switched systems is not in general obtainable, automated synthesis for this type of complex systems is naturally very challenging especially with respect to high-level logic properties, \emph{e.g.,} linear temporal logic (LTL) formulae~\cite{pnueli1977temporal}.

To alleviate the encountered computational complexity,  approximation techniques have been proposed in relevant literatures, where original dynamics are approximated by simpler ones with finite state sets, \emph{i.e.,} finite abstractions. However, one major bottleneck in the existing approximation techniques is the state-explosion problem. To tackle this issue, some recent works (\emph{e.g.,}~\cite{lavaei2017compositional,lavaei2018ADHS,lavaei2017HSCC,lavaei2019LSS,lavaei2019ECC,lavaei2019CDC,lavaei2018CDCJ,lavaei2019HSCC_J,lavaei2018ADHSJ,lavaei2019NAHS,lavaei2019NAHSJ,lavaei2019Thesis,AmyIFAC2020,AmyJournal2020}) study different compositional schemes for the construction of (in)finite abstractions for complex stochastic systems via (in)finite abstractions of their smaller subsystems.

\emph{Control barrier certificates}, as a discretization-free approach for controller synthesis of complex systems, have been introduced in recent years as another potential solution to mitigate the computational complexity arising in the analysis or synthesis of large-scale stochastic systems. In this respect,  discretization-free techniques based on barrier certificates for stochastic hybrid systems are initially proposed in~\cite{prajna2007framework}. Stochastic safety verification using barrier certificates for switched diffusion processes and stochastic hybrid systems is, respectively, proposed in~\cite{wisniewski2017stochastic} and~\cite{huang2017probabilistic}. A verification approach for stochastic switched systems against safe LTL objectives via barrier functions is proposed in~\cite{Mahathi2019}. Temporal logic verification of stochastic systems via control barrier certificates and its extension to formal synthesis are, respectively, proposed in~\cite{jagtap2018temporal} and~\cite{Pushpak2019}. Recently, compositional construction of control barrier certificates for large-scale stochastic discrete-time and continuous-time systems is respectively presented in~\cite{LavaeiIFAC2020} and~\cite{AmyIFAC12020}.

It should be noted that although~\cite{lavaei2019HSCC_J} provides a compositional approach for the same class of stochastic switched system as in this work, their proposed framework is based on the construction of finite abstractions which relies on the discretization of state and input sets and consequently suffers from the curse of dimensionality problem. In contrast, we propose here a compositional framework, for the first time for stochastic switched systems, based on control barrier certificates.

{\bf Contributions.} 
In this paper, we propose a compositional framework for the construction of control barrier certificates for large-scale stochastic switched systems accepting \emph{multiple} control barrier certificates with some \emph{dwell-time} conditions. To this end, we first provide an augmented framework for presenting
each switched subsystem with several modes with a single system covering all modes (called augmented switched systems) whose output trajectories are exactly the same as those of original switched systems. We then compositionally construct augmented control barrier certificates for interconnected augmented systems based on so-called \emph{augmented pseudo-barrier certificates} of subsystems by leveraging some $\max$-type small-gain conditions. Afterwards, given the constructed augmented barrier certificates, we quantify upper bounds on the probability that interconnected systems reach certain unsafe regions in a finite-time horizon. We finally utilize a technique based on a counter-example guided inductive synthesis (CEGIS) approach to search for control barrier certificates of each mode. 

To illustrate the effectiveness of our proposed results, we first apply them to a room temperature network in a circular building containing $1000$ rooms and compositionally synthesize safety controllers to keep the temperature of each room in a comfort zone in a bounded time horizon. Eventually, to show the applicability of our results to switched systems accepting \emph{multiple} barrier certificates with a \emph{dwell-time} condition, we apply our proposed techniques to a circular cascade network of $500$ subsystems (totally $1000$ dimensions) and provide upper bounds on the probability that the interconnected system reaches some unsafe region in a finite-time horizon. 

{\bf Related Work.} Although the proposed results in~\cite{wisniewski2017stochastic,huang2017probabilistic} deal with stochastic switched and a class of hybrid systems, their ultimate goal is to perform probabilistic safety verification via barrier certificates in a monolithic manner. In comparison, in this work, we propose a compositional framework for the construction of control barrier certificates for large-scale stochastic switched systems admitting \emph{multiple} barrier certificates with some \emph{dwell-time} conditions. We utilize those barrier certificates and conditions to compositionally synthesize state-feedback controllers for interconnected systems enforcing safety specifications over a finite-time horizon.

\section{Discrete-Time Stochastic Switched Systems}

\subsection{Preliminaries}
We consider a probability space $(\Omega,\mathcal F_{\Omega},\mathds{P}_{\Omega})$,
where $\Omega$ is the sample space,
$\mathcal F_{\Omega}$ is a sigma-algebra on $\Omega$ comprising subsets of $\Omega$ as events,
and $\mathds{P}_{\Omega}$ is a probability measure that assigns probabilities to events.
We assume that random variables introduced in this article are measurable functions of the form $X:(\Omega,\mathcal F_{\Omega})\rightarrow (S_X,\mathcal F_X)$.
Any random variable $X$ induces a probability measure on  its space $(S_X,\mathcal F_X)$ as $Prob\{A\} = \mathds{P}_{\Omega}\{X^{-1}(A)\}$ for any $A\in \mathcal F_X$.
We often directly discuss the probability measure on $(S_X,\mathcal F_X)$ without explicitly mentioning the underlying probability space and the function $X$ itself.

A topological space $S$ is called a Borel space if it is homeomorphic to a Borel subset of a Polish space (\emph{i.e.,} a separable and completely metrizable space).
Examples of a Borel space are the Euclidean spaces $\mathbb R^n$, its Borel subsets endowed with a subspace topology, as well as hybrid spaces.
Any Borel space $S$ is assumed to be endowed with a Borel sigma-algebra, which is
denoted by $\mathds B(S)$. We say that a map $f : S\rightarrow Y$ is measurable whenever it is Borel measurable.

\subsection{Notation}

The following notation is employed throughout the paper. We denote the set of real, positive and non-negative real numbers by $\mathbb{R},\mathbb{R}_{>0}$, and $\mathbb{R}_{\geq 0}$, respectively. We use $\mathbb{R}^n$ to denote a real space of $n$ dimension. $\mathbb{N} := \{0,1,2,...\}$ represents the set of non-negative integers and $\mathbb{N}_{\geq 1}=\{1,2,...\}$ is the set of positive integers. Given $N$ vectors $x_i \in \mathbb{R}^{n_i}$, $x=[x_1;...;x_N]$ denotes the corresponding vector of dimension $\sum_i n_i$. Given a vector $x\in\mathbb{R}^{n}$, $\Vert x\Vert$ denotes the infinity norm of $x$. Symbols $\mathds{I}_n$, $\mathbf{0}_n$, and $\mathds{1}_n$ denote the identity matrix in $\mathbb R^{n\times{n}}$ and the column vector in $\mathbb R^{n\times{1}}$ with all elements equal to zero and one, respectively. The identity
function and composition of functions are denoted by $\mathcal{I}_d$ and symbol $\circ$, respectively. Given functions $f_i:X_i\rightarrow Y_i$, for any $i\in\{1,\ldots,N\}$, their Cartesian product $\prod_{i=1}^{N}f_i:\prod_{i=1}^{N}X_i\rightarrow\prod_{i=1}^{N}Y_i$ is defined as $(\prod_{i=1}^{N}f_i)(x_1,\ldots,x_N)=[f_1(x_1);\ldots;f_N(x_N)]$.  A function $\varphi: \mathbb{R}_{\geq 0} \rightarrow \mathbb{R}_{\geq 0}$ is said to be a class $\mathcal{K}$ function if it is continuous, strictly increasing, and $\varphi(0)=0$. A class $\mathcal{K}$ function $\varphi(\cdot)$ belongs to class $\mathcal{K}_\infty$ if $\varphi(s) \rightarrow \infty$ as $s \rightarrow \infty$. 

\subsection{Discrete-Time Stochastic Switched Systems}
We consider stochastic switched systems in discrete-time (dt-SS) defined formally as follows.
\begin{definition}
	A discrete-time stochastic switched system (dt-SS) is characterized by the tuple
	\begin{equation}
		\label{eq:dt-SS}
		\Sigma=(X,P,\mathcal{P},W,\varsigma, F, Y, h),
	\end{equation}
	where: 
	\begin{itemize}
		\item $X\subseteq \mathbb R^n$ is a Borel space as the state set of the system. We denote by $(X, \mathds B (X))$ the measurable space with $\mathds B (X)$  being  the Borel sigma-algebra on the state space;
		\item $P = \{1,\dots, m \}$  is a finite set of modes;
		\item $\mathcal{P}$ is a subset of $\mathcal{S}(\mathbb N,P)$ which denotes the set of functions from $\mathbb N$ to $P$;
		\item $W\subseteq \mathbb R^{\bar p}$ is a Borel space as the \emph{internal} input set of the system; 
		\item $\varsigma$ is a sequence of independent and identically distributed (i.i.d.) random variables on a set $V_\varsigma$
		\begin{equation*}
			\varsigma:=\{\varsigma(k):\Omega\rightarrow V_{\varsigma},\,\,k\in\N\};
		\end{equation*}
		\item $F = \{f_1,\dots, f_m \}$ is a collection of vector fields indexed by $p$. For all $p\in P$, the map $f_p:X\times W\times V_{\varsigma} \rightarrow X$ is a measurable function characterizing the state evolution of the system in mode $p$;
		\item  $Y\subseteq \mathbb R^{q}$ is a Borel space as the output set of the system;
		\item  $h:X\rightarrow Y$ is a measurable function as the output map that maps a state $x\in X$ to its output $y = h(x)$.
	\end{itemize}
	
	For a given initial state $x(0)\in X$, an internal input sequence $w(\cdot):\mathbb N\rightarrow W$, and a switching signal $\bold{p}(k):\mathbb N \rightarrow P$, the evolution of the state of $\Sigma$ is described as
	\begin{equation}\label{Eq_1a}
		\Sigma:\left\{\hspace{-1.5mm}\begin{array}{l}x(k+1)=f_{\bold{p}(k)}(x(k),w(k),\varsigma(k)),\\
			y(k)=h(x(k)),\\
		\end{array}\right.
		\quad\quad k\in\mathbb N.
	\end{equation}
\end{definition}
We assume that the signal $\bold{p}$ satisfies a \emph{dwell-time} condition~\cite{morse1996supervisory} as defined in the next definition.
\begin{definition}\label{dwell-time}
	Consider a switching signal $\bold{p}:\mathbb N\rightarrow P$ and define its switching time instants as
	$$\mathfrak S_{\bold{p}} := \left\{\mathfrak s_k: k\in\mathbb N_{\ge 1}\right\}\!.$$ Then, $\bold{p}:\mathbb N\rightarrow P$ has \emph{dwell-time} $k_d \in \mathbb N$~\cite{morse1996supervisory} if elements of $\mathfrak S_{\bold{p}}$ ordered as $\mathfrak s_1\le \mathfrak s_2\le \mathfrak s_3\le \dots$ satisfy $\mathfrak s_1\geq k_d$ and $\mathfrak s_{k+1} - \mathfrak s_{k} \geq k_d, \forall k\in\mathbb N_{\ge 1}$.
\end{definition}

	For any $p\in P$, we use $\Sigma_p$ to refer to system~\eqref{Eq_1a} with a constant switching signal $\bold{p}(k) = p$ for all $k\in\mathbb N$. We are ultimately interested in investigating interconnected dt-SS without internal inputs that result from the interconnection of dt-SS having both internal and external inputs. In this case, the interconnected dt-SS without internal inputs is indicated by the simplified tuple $(X,P, \mathcal{P},\varsigma, F, Y, h)$ where $f_p:X\times V_\varsigma\rightarrow X$, $\forall p\in P$.

\subsection{Augmented Stochastic Switched Systems}

Here, given a dt-SS $\Sigma$, we introduce the notion of augmented dt-SS as in the next definition. Note that this notion is adapted from the definition of labeled transition systems defined in~\cite{baier2008principles} and modified to capture the stochastic nature of the system. This provides an alternative description of switched systems enabling us to represent a switched system with a finite set of modes via an augmented system covering the whole modes. 

\begin{definition}\label{def: ASS}
	Given a dt-SS $\Sigma=(X,P,\mathcal{P},W,\varsigma,F,Y,h)$, we define the associated augmented dt-SS $\mathbb{A}(\Sigma) = (\mathbb{X},\mathbb{P},\mathbb{W},\varsigma,\mathbb{F},\mathbb{Y},\mathbb{H})$,
	where:
	\begin{itemize}
		\item $\mathbb{X} = X \times P \times \{0,\dots,k_d-1\}$ is the set of states. A state $(x,p,l) \in \mathbb{X}$ means that the current state of $\Sigma$ is $x$, the current value of the switching signal is $p$, and the time elapsed since the latest switching time instant upper bounded by $k_d$ is $l$;
		\item $\mathbb{P} = P$ is the set of \emph{external} inputs;
		\item $\mathbb{W}=W$ is the set of \emph{internal} inputs;
		\item $\varsigma$ is a sequence of i.i.d. random variables;
		\item $\mathbb{F}:\mathbb{X}\times \mathbb{P}\times\mathbb{W}\times V_{\varsigma} \rightarrow \mathbb{X}$\, is the one-step transition function given by $(x',p',l')=\mathbb{F}\,((x,p,l),p,w,\varsigma)$ if and only if $x'=f_p(x,w,\varsigma)$ and the following scenarios hold:
		\begin{itemize}
			\item $l<k_d-1, p'=p$, and $l' = l+1$: switching is not allowed because the time elapsed since the latest switch is strictly smaller than the dwell-time;
			\item $l=k_d-1, p'=p$, and $l'=k_d-1$: switching is allowed but no switch occurs;
			\item $l=k_d-1, p'\neq p$, and $l' = 0$: switching is allowed and a switch occurs;
		\end{itemize}	
		\item $\mathbb{Y} = Y$ is the output set;
		\item $\mathbb{H}:\mathbb{X} \rightarrow \mathbb{Y}$ is the output map defined as $\mathbb{H}\,(x,p,l) = h(x)$.
	\end{itemize}
	We associate respectively to $\mathbb{P}$ and $\mathbb{W}$ the sets $\mathsf P$ and $\mathsf W$ to be collections of sequences $\{p(k):\Omega\rightarrow \mathbb{P},\,\,k\in\N\}$ and $\{w(k):\Omega\rightarrow \mathbb{W},\,\,k\in\N\}$, in which $p(k)$ and $w(k)$ are independent of $\varsigma(t)$ for any $k,t\in\mathbb N$ and $t\ge k$. We also denote initial conditions of $p$ and $l$ by $p_0$ and $l_0=0$.	
\end{definition}

\begin{remark}
	Note that in the augmented dt-SS $\mathbb{A}(\Sigma)$ in Definition~\ref{def: ASS}, we added two additional variables $p$ and $l$ to the state tuple of the system $\Sigma$, in which $l$ is a counter that depending on its value allows or prevents the system from switching, and $p$ acts as a memory to record the latest mode.
\end{remark}

\begin{proposition}\label{Proposition}
	The output trajectory of the augmented dt-SS $\mathbb{A}(\Sigma)$ in Definition~\ref{def: ASS} can be uniquely mapped to an output trajectory of the switched system $\Sigma$ defined in~\eqref{Eq_1a}, and vice versa.
\end{proposition}

The proof is similar to that of~\cite[Proposition 2.9]{lavaei2019HSCC_J} and is
omitted here.

In the next section, in order to quantify upper bounds on the probability that
the interconnected system reaches a certain unsafe region
in a finite-time horizon, we first introduce notions of augmented control pseudo-barrier and barrier certificates for, respectively, augmented dt-SS (with both internal and external signals) and interconnected augmented dt-SS (without internal signals).

\section{Augmented Control (Pseudo-)Barrier Certificates}\label{sec:SPSF}

Here, we first introduce a notion of augmented control pseudo-barrier certificates for augmented dt-SS with both internal and external inputs.
\begin{definition}\label{Def_1a} 
	Consider an augmented dt-SS $\mathbb{A}(\Sigma) = (\mathbb{X},\mathbb{P},\mathbb{W},\varsigma,\mathbb{F},\mathbb{Y},\mathbb{H})$, and initial and unsafe sets $X_0, X_1 \subseteq  X$ for the dt-SS $\Sigma$. Let us define $\mathbb{X}_0 = X_0 \times P \times \{0\},~\mathbb{X}_1 = X_1 \times P \times \{0,\dots,k_d-1\}$, as initial and unsafe sets of the augmented system, respectively.
	A function $\mathcal B:\mathbb{X}\to\R_{\ge0}$ is called an augmented control
	pseudo-barrier certificate (APBC) for $\mathbb{A}(\Sigma)$ if there exist functions $\alpha\in\mathcal{K}_\infty$, $\rho_{\mathrm{int}}\in\mathcal{K}_\infty\cup\{0\}$, and constants $0<\kappa<1$, $\gamma,\psi\in\R_{\geq 0}$ and $\lambda\in\R_{> 0}$, such that
	\begin{align}\label{Eq_2a}
		&\mathcal B(x,p,l) \geq \alpha(\Vert \mathbb{H}(x,p,l)\Vert),\quad\quad \forall(x,p,l)\in \mathbb{X},\\\label{Eq_2a1}
		&\mathcal B(x,p,l) \leq \gamma,\quad\quad\quad\quad\quad\quad\quad\quad\!\! \forall(x,p,l)\in \mathbb{X}_0,\\\label{Eq_2a2}
		&\mathcal B(x,p,l) \geq \lambda, \quad\quad\quad\quad\quad\quad\quad\quad\!\! \forall(x,p,l)\in \mathbb{X}_1, 
	\end{align}  
	and $\forall(x,p,l)\in \mathbb{X}$, $\exists p'\in \mathbb{P}$, such that $\forall w\in \mathbb{W}$, one has $(x',p',l')=\mathbb{F}\,((x,p,l),p,w,\varsigma)$, and
	\begin{align}\label{Eq_3a}
		\mathbb{E} \Big[\mathcal B((x',p',l'))\,\big|\,x,p,l,w\Big]\leq \max\Big\{\kappa \mathcal B(x,p,l), \rho_{\mathrm{int}}(\Vert w\Vert),\psi\Big\},
	\end{align}
	where the expectation operator $\mathbb E$ is with respect to $\varsigma$ under the one-step transition of the augmented dt-SS $\mathbb{A}(\Sigma)$.
\end{definition}

Now, we modify the above notion for augmented dt-SS without internal inputs by eliminating all the terms related to $w$ which will be employed later for relating interconnected augmented switched systems.

\begin{definition}\label{Def_2a} 
	Consider an (interconnected) augmented dt-SS $\mathbb{A}(\Sigma) = (\mathbb{X},\mathbb{P},\varsigma,\mathbb{F},\mathbb{Y},\mathbb{H})$ without internal inputs, with initial and unsafe sets $X_0, X_1 \subseteq  X$ for the dt-SS $\Sigma$. Let us define sets $\mathbb X_0, \mathbb X_1 \subseteq \mathbb X$ as respectively initial and unsafe sets of the augmented system.
	A function $\mathcal B:\mathbb{X}\to\R_{\ge0}$ is called an augmented control barrier certificate (ABC) for $\mathbb{A}(\Sigma)$ if
	\begin{align}\label{Eq_2}
		&\mathcal B(x,p,l) \leq \gamma,\quad\quad\quad\quad\quad \forall(x,p,l)\in \mathbb{X}_0,\\\label{Eq_3}
		&\mathcal B(x,p,l) \geq \lambda,\quad\quad\quad\quad\quad \forall(x,p,l)\in \mathbb{X}_1,
	\end{align}  
	and $\forall(x,p,l)\in \mathbb{X}, \exists p'\in \mathbb{P}$, such that one has $(x',p',l')=\mathbb{F}\,((x,p,l),p,\varsigma)$, and
	\begin{align}\label{eq6666}
		\mathbb{E} \Big[&\mathcal B((x',p',l'))\,\big|\,x,p,l\Big]\leq \max\Big\{\kappa \mathcal B(x,p,l),\psi\Big\},\
	\end{align}
	for some constants $0<\kappa<1$, $\gamma,\psi\in\R_{\geq 0}$ and $\lambda\in\R_{> 0}$ with $\gamma < \lambda$, where the expectation operator $\mathbb E$ is with respect to $\varsigma$ under the one-step transition of the augmented dt-SS $\mathbb{A}(\Sigma)$.
\end{definition}

Now we employ Definition~\ref{Def_2a} and propose an upper bound on the probability that an (interconnected) augmented dt-SS reaches an unsafe region via the next theorem.

\begin{theorem}\label{Kushner}
	Let $\mathbb{A}(\Sigma) = (\mathbb{X},\mathbb{P},\varsigma,\mathbb{F},\mathbb{Y},\mathbb{H})$ be an (interconnected) augmented dt-SS without internal inputs. Suppose $\mathcal B$ is an ABC for $\mathbb{A}(\Sigma)$. Then for any random variable $a$ as the initial state, any initial mode $p_0$, and $l_0 = 0$ as the initial counter, the probability that the interconnected augmented dt-SS reaches an unsafe set $\mathbb X_1$ within the time step $k\in [0,T_d]$ is upper bounded by $\delta$ as
	\begin{equation}\label{eqlemma2}
		\mathbb{P} \Big\{\sup_{0 \leq k \leq T_d} \mathcal B(x(k),p(k),l(k)) \geq \lambda \,\, \big|\,\, a,p_0,l_0\Big\} \leq \delta,
	\end{equation}
	where	
	\begin{equation*}
		\delta=  \begin{cases} 
			1-(1-\frac{\gamma}{\lambda})(1-\frac{\psi}{\lambda})^{T_d}, & \quad\quad \text{if } \lambda \geq \frac{\psi}{{\kappa}}, \\
			(\frac{\gamma}{\lambda})(1-{\kappa})^{T_d}+(\frac{\psi}{{\kappa}\lambda})(1-(1-{\kappa})^{T_d}), & \quad\quad\text{if } \lambda< \frac{\psi}{{\kappa}}.  \\
		\end{cases}
	\end{equation*}	
\end{theorem}

The proof of Theorem~\ref{Kushner} is provided in Appendix.

\section{Compositional Construction of ABC}\label{sec:compositionality}
In this section, we analyze networks of stochastic switched subsystems by driving a $\max$-type small-gain condition and discuss how to construct an ABC of the augmented dt-SS via the corresponding APBC of subsystems. 

Suppose we are given $N$ stochastic switched subsystems
\begin{equation}
	\label{eq:network}
	\Sigma_i=(X_i,P_i, \mathcal{P}_i,W_i,\varsigma_i, F_i, Y_i, h_i),~i\in \{1,\dots,N\},
\end{equation}
where $F_i=\{f_1^i,\ldots,f_{m_i}^i\}$, with its \emph{equivalent} augmented dt-SS $\mathbb{A}(\Sigma_i) = (\mathbb{X}_i,\mathbb{P}_i,\mathbb{W}_i,\varsigma_i,\mathbb{F}_i,\mathbb{Y}_i,\mathbb{H}_i)$, in which their internal inputs and outputs are partitioned as
\begin{align}\label{config1}
	w_i=[{w_{i1};\ldots;w_{i(i-1)};w_{i(i+1)};\ldots;w_{iN}}],\quad y_i=[{y_{i1};\ldots;y_{iN}}],
\end{align}
and their output sets and functions are of the form
\begin{align}\label{config2}
	Y_i=\prod_{j=1}^{N}Y_{ij},\quad h_{i}(x_i)=[h_{i1}(x_i);\ldots;h_{iN}(x_i)].
\end{align}
We interpret outputs $y_{ii}$ as \emph{external} ones, whereas outputs $y_{ij}$ with $i\neq j$ are \emph{internal} ones which are utilized to interconnect these stochastic switched subsystems. For the interconnection, we assume that $w_{ij}$ is equal to $y_{ji}$ if there is a connection from $\Sigma_{j}$ to $\Sigma_i$, otherwise we put the connecting output function identically zero, \emph{i.e.,} $h_{ji}\equiv 0$.

Now, we are ready to define the \emph{interconnection} of dt-SS $\Sigma_i=(X_i,P_i, \mathcal{P}_i,W_i,\varsigma_i, F_i, Y_i, h_i)$.
\begin{definition}
	Consider $N\in\mathbb N_{\geq1}$ dt-SS $\Sigma_i=(X_i,P_i, \mathcal{P}_i,W_i,\varsigma_i, F_i, Y_i, h_i)$, with the input-output configuration as in \eqref{config1}-\eqref{config2}. The interconnection of  $\Sigma_i$, $\forall i\in \{1,\ldots,N\}$, is the interconnected dt-SS $\Sigma = (X,P, \mathcal{P},\varsigma, F, Y, h)$, denoted by
	$\mathcal{I}(\Sigma_1,\ldots,\Sigma_N)$, such that $X:=\prod_{i=1}^{N}X_i$,  $P:=\prod_{i=1}^{N}P_i$, $\mathcal{P}:=\prod_{i=1}^{N}\mathcal{P}_i$, $F:=\prod_{i=1}^{N}F_{i}$, $Y:=\prod_{i=1}^{N}Y_{ii}$, and $h=\prod_{i=1}^{N}h_{ii}$, subjected to the following constraint:
	\begin{equation*}
		\forall i,j\in \{1,\dots,N\},i\neq j\!: \quad\quad w_{ji} = y_{ij}, \quad Y_{ij}\subseteq W_{ji}.
	\end{equation*}
	
\end{definition}

\begin{figure}[ht]
	\begin{tikzpicture}[>=latex']
	\tikzstyle{block} = [draw, 
	thick,
	rectangle, 
	minimum height=.8cm, 
	minimum width=1.5cm]
	
	\node at (-3.5,-0.75) {$\mathcal{I}(\Sigma_1,\Sigma_2)$};
	
	\draw[dashed] (-1.7,-2.2) rectangle (1.7,.7);
	
	\node[block] (S1) at (0,0) {$\Sigma_1$};
	\node[block] (S2) at (0,-1.5) {$\Sigma_2$};
	
	\draw[->] ($(S1.east)+(0,0.25)$) -- node[very near end,above] {$y_{11}$} ($(S1.east)+(1.5,.25)$);
	\draw[<-] ($(S1.west)+(0,0.25)$) -- node[very near end,above] {$\nu_{1}$} ($(S1.west)+(-1.5,.25)$);
	
	\draw[->] ($(S2.east)+(0,-.25)$) -- node[very near end,below] {$y_{22}$} ($(S2.east)+(1.5,-.25)$);
	\draw[<-] ($(S2.west)+(0,-.25)$) -- node[very near end,below] {$\nu_{2}$} ($(S2.west)+(-1.5,-.25)$);
	
	\draw[->] 
	($(S1.east)+(0,-.25)$) -- node[very near end,above] {$y_{12}$} 
	($(S1.east)+(.5,-.25)$) --
	($(S1.east)+(.5,-.5)$) --
	($(S2.west)+(-.5,.5)$) --
	($(S2.west)+(-.5,.25)$) -- node[very near start,below] {$w_{21}$}
	($(S2.west)+(0,.25)$) ;
	
	\draw[->] 
	($(S2.east)+(0,.25)$) -- node[very near end,below] {$y_{21}$} 
	($(S2.east)+(.5,.25)$) --
	($(S2.east)+(.5,.5)$) --
	($(S1.west)+(-.5,-.5)$) --
	($(S1.west)+(-.5,-.25)$) -- node[very near start,above] {$w_{12}$}
	($(S1.west)+(0,-.25)$) ;
	
	\end{tikzpicture}
	\caption{An interconnection of two stochastic subsystems $\Sigma_1$ and $\Sigma_2$.}
	\label{system1}
\end{figure}
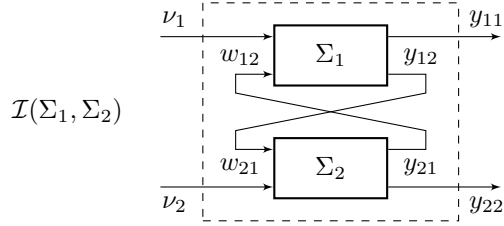

An example of the interconnection of two stochastic subsystems $\Sigma_1$ and $\Sigma_2$ is illustrated in Figure \ref{system1}.

Similarly, we define a notion of the \emph{interconnection} for augmented dt-SS $\mathbb{A}(\Sigma_i) = (\mathbb{X}_i,\mathbb{P}_i,\mathbb{ W}_i,\varsigma_i,\mathbb{ F}_i,\mathbb{Y}_i,\mathbb{H}_i)$.

\begin{definition}
	Consider $N\in\mathbb N_{\geq1}$ augmented dt-SS $\mathbb{A}(\Sigma_i) = (\mathbb{X}_i,\mathbb{P}_i,\mathbb{W}_i,\varsigma_i,\mathbb{F}_i,\mathbb{Y}_i,\mathbb{H}_i)$, with the input-output configuration as in~\eqref{config1}-\eqref{config2}. The interconnection of  $\mathbb{A}(\Sigma_i)$, $\forall i\in \{1,\ldots,N\}$, is the interconnected augmented dt-SS $\mathbb{A}(\Sigma) = (\mathbb{X},\mathbb{P},\varsigma,\mathbb{F},\mathbb{Y},\mathbb{H})$, denoted by
	$\mathcal{I}(\mathbb{A}(\Sigma_1),\ldots,\mathbb{A}(\Sigma_N))$, such that $\mathbb{X}:=\prod_{i=1}^{N}\mathbb{X}_i$,  $\mathbb{P}:=\prod_{i=1}^{N}\mathbb{P}_i$, $\mathbb{Y}:=\prod_{i=1}^{N} \mathbb{Y}_{ii}$, $\mathbb{H}=\prod_{i=1}^{N}\mathbb{H}_{ii}$, and the map $\mathbb{F}=\prod_{i=1}^{N}\mathbb{F}_{i}$  is the transition function given by $(x',p',l')=\mathbb{F}\,((x,p,l),p,\varsigma)$ if and only if $ x' = f_p({x},{w},\varsigma)$, where $f_p=\prod_{i=1}^{N}f_{p_i}^i$, and the following scenarios hold for any $i\in\{1,\dots,N\}$: 
	\begin{itemize}
		\item $l_i<k_{d_i}-1$, $p'_i = p_i$, and $l'_i = l_i+1$;
		\item $l_i=k_{d_i}-1$, $p'_i = p_i$, and $l'_i=k_{d_i}-1$;
		\item $l_i=k_{d_i}-1$, $p'_i\neq p_i$, and $l'_i = 0$;
	\end{itemize}where $x= [x_1;\dots;x_N], p= [p_1;\dots;p_N], l= [l_1;\dots;l_N], \varsigma= [\varsigma_1;\dots;\varsigma_N],$
	and subjected to the following constraint:
	\begin{align}\notag
		\forall i,j\in \{1,\dots,N\},i\neq j\!: \quad\quad w_{ji} = y_{ij},\quad \mathbb{Y}_{ij}\subseteq\mathbb{W}_{ji}.
	\end{align}
\end{definition}

Assume for the augmented dt-SS $\mathbb{A}(\Sigma_i) = (\mathbb{X}_i,\mathbb{P}_i,\mathbb{W}_i,\varsigma_i,\mathbb{F}_i,\mathbb{Y}_i,\mathbb{H}_i),i\in \{1,\dots,N\}$, there exists an APBC $\mathcal B_i$ with the corresponding functions and constants denoted by $\alpha_i, \rho_{\mathrm{int}i},\kappa_i, \gamma_i, \lambda_i$ and $\psi_i$ as in Definition~\ref{Def_1a}. Now we raise the following $\max$ small-gain assumption to establish the main compositionality result of the paper.

\begin{assumption}\label{Assump: Gamma}
	Assume that $\mathcal{K}_\infty$ functions $\kappa_{ij}$ defined as
	\begin{equation*}
		\kappa_{ij}(s) := 
		\begin{cases}
			\kappa_is, \quad\quad& \text{if }i = j,\\
			\rho_{\mathrm{int}i}(\alpha_j^{-1}(s)), \quad\quad& \text{if }i \neq j,
		\end{cases}
	\end{equation*}
	satisfy
	\begin{equation}\label{Assump: Kappa}
		\kappa_{i_1i_2}\circ\kappa_{i_2i_3}\circ\dots \circ \kappa_{i_{r-1}i_{r}}\circ\kappa_{i_{r}i_1} < \mathcal{I}_d,
	\end{equation}	
	for all sequences $(i_1,\dots,i_{r}) \in \{1,\dots,N\}^ {r}$ and $r\in \{1,\dots,N\}$. 
\end{assumption}
The small-gain condition~\eqref{Assump: Kappa} implies the existence of $\mathcal{K}_\infty$ functions $\sigma_i>0$ \cite[Theorem 5.5]{ruffer2010monotone}, satisfying
\begin{align}\label{compositionality}
	\max_{i,j}\Big\{\sigma_i^{-1}\circ\kappa_{ij}\circ\sigma_j\Big\} < \mathcal{I}_d, ~~~~i,j = \{1,\dots,N\}.
\end{align}
\begin{remark}
	Note that the small-gain condition~\eqref{Assump: Kappa} is a standard one in studying the stability of large-scale interconnected systems via ISS Lyapunov functions~\cite{dashkovskiy2007iss,dashkovskiy2010small}. This condition is automatically satisfied if each $\kappa_{ij}$
	is less than identity (i.e., $\kappa_{ij}<\mathcal{I}_d, \forall i,j\in\{1,\dots,N\}$).
\end{remark}		
In the next theorem, we show that if Assumption \ref{Assump: Gamma} holds and $\max_{i}\sigma_i^{-1}$ is concave (in order to employ Jensen's inequality), then we can construct an ABC of $\mathbb{A}(\Sigma)$ using the APBC of $\mathbb{A}(\Sigma_i)$.

\begin{theorem}\label{Thm: Comp}
	Consider the interconnected augmented dt-SS $\mathbb{A}(\Sigma) = (\mathbb{X},\mathbb{P},\varsigma,\mathbb{F},\mathbb{Y},\mathbb{H})$ induced by $N\in\mathbb N_{\geq1}$ augmented dt-SS $\mathbb{A}(\Sigma_i)$. Suppose that each $\mathbb{A}(\Sigma_i)$ admits an APBC $\mathcal B_i$ as defined in Definition~\ref{Def_1a}. If Assumption~\ref{Assump: Gamma} holds and
	\begin{align}\label{compositionality1}
		\max_{i} \Big\{\sigma_i^{-1}(\lambda_i)\Big\} > \max_{i} \Big\{\sigma_i^{-1}(\gamma_i)\Big\},
	\end{align}
	then the function $\mathcal B(x,p,l)$ defined as
	\begin{equation}\label{Comp: Simulation Function}
		\mathcal B(x,p,l) := \max_{i} \Big\{\sigma_i^{-1}(\mathcal B_i(x_i,p_i,l_i))\Big\},
	\end{equation}
	is an ABC for the interconnected augmented dt-SS $\mathcal{I}(\mathbb{A}(\Sigma_1),\ldots,\mathbb{A}(\Sigma_N))$ provided that $\max_{i}\sigma_i^{-1}$ for $\sigma_i$ as in \eqref{compositionality} is concave. 	
\end{theorem}

The proof of Theorem~\ref{Thm: Comp} is provided in Appendix.

\begin{remark}
	Note that the condition~\eqref{compositionality1} in general is not very restrictive since functions $\sigma_{i}$ in~\eqref{compositionality} play an important role in rescaling APBC for subsystems while normalizing the effect of internal gains of other subsystems (cf. \cite{dashkovskiy2010small} for a similar argument but in the context of stability analysis via ISS Lyapunov functions). Then one can expect that the condition~\eqref{compositionality1} holds in many applications due to this rescaling.
\end{remark}

\section{Construction of APBC}\label{sec:constrcution_finite}

In this section, we impose conditions on the dt-SS $\Sigma_p$ enabling us to find an APBC for $\mathbb{A}(\Sigma)$. The APBC for the augmented dt-SS $\mathbb{A}(\Sigma)$ is established under the assumption that the given dt-SS $\Sigma_p$ has control barrier certificates (CBC) for all modes as in the following definition.

\begin{definition}\label{cbc}
	Consider a dt-SS $\Sigma_p$, and sets $X_0, X_1 \subseteq X$ as initial and unsafe sets of the given dt-SS, respectively. A function $\mathcal B_p:X \rightarrow \mathbb{R}_{\geq 0}$ is said to be a control barrier certificate (CBC) for $\Sigma_p$ if there exist functions
	$\alpha_p\in\mathcal{K}_\infty$, $\rho_{\mathrm{int}p}\in\mathcal{K}_\infty\cup\{0\}$, and constants $0<\kappa_p<1$, $\gamma_p,\psi_p\in\R_{\geq 0}$ and $\lambda_p\in\R_{> 0}$, such that
	\begin{align}\label{subsys1}
		&\mathcal B_p(x) \geq \alpha_p(\Vert h(x)\Vert),\quad\quad\forall x \in X,\\\label{subsys2}
		&\mathcal B_p(x) \leq \gamma_p,\quad\quad\quad\quad\quad\quad\!\forall x \in X_{0},\\\label{subsys3}
		&\mathcal B_p(x) \geq \lambda_p, \quad\quad\quad\quad\quad\quad\!\forall x \in X_{1}, 
	\end{align}  
	and $\forall x\in X$, $\forall w\in W$, one has
	\begin{align}\label{csbceq}
		\mathbb{E}\Big[\mathcal B_p(x(k+1)) \,\big|\, x, w\Big]\leq \max\Big\{\kappa_p\mathcal B_p(x), \rho_{\mathrm{int}p}(\|w\|),\psi_p\Big\}.
	\end{align}
\end{definition}

In order to construct an APBC for the augmented dt-SS $\mathbb{A}(\Sigma)$, we need also to raise the following assumption.

\begin{assumption}\label{Assume: Switching}
	Suppose there exists $\mu\ge 1$ such that
	\begin{align}\label{Barrier_rate}
		\forall x \in X,~ \forall p,p' \in P, \quad \mathcal B_p(x)\leq \mu \mathcal B_{p'}(x).
	\end{align}
\end{assumption}

\begin{remark}
	Assumption~\ref{Assume: Switching} is a standard one in the literature for switched systems accepting \emph{multiple} Lyapunov functions with \emph{dwell-time} similar to the one appeared in~\cite[equation (3.6)]{liberzon2003switching}. 
\end{remark}

Under Definition~\ref{cbc} and Assumption~\ref{Assume: Switching}, the next theorem lays the foundations for constructing an APBC for $\mathbb{A}(\Sigma)$.

\begin{theorem}\label{Thm_5a}
	Let $\Sigma=(X,P, \mathcal{P},W,\varsigma, F, Y, h)$ be a switched subsystem with its equivalent augmented system $\mathbb{A}(\Sigma) = (\mathbb{X},\mathbb{P},\mathbb{W},\varsigma,\mathbb{F},\mathbb{Y},\mathbb{H})$. Let $\mathcal B_p$ be a CBC for  $\Sigma_p$, $\forall p\in P$, as in Definition~\ref{cbc}, and assume Assumption~\ref{Assume: Switching} holds, and consider $\epsilon>1$. If \,$\forall p\in P$, $~k_d \geq \epsilon\frac{\ln(\mu)}{\ln(1/\kappa_p)}+1$, then 
	\begin{align}\label{function V}
		\mathcal B(x,p,l)=\frac{1}{{\kappa_p}^{l/\epsilon}}\mathcal B_p(x),
	\end{align} 
	is an APBC for $\mathbb{A}(\Sigma)$.
\end{theorem}

The proof of Theorem~\ref{Thm_5a} is provided in Appendix.

\begin{remark}\label{Common_barrier}
	Note that if there exists a common CBC $\mathcal B: X\to\R_{\ge0}$ for all switching modes $p\in P$ satisfying conditions of Definition~\ref{cbc} and Assumption~\ref{Assume: Switching} (with $\mu = 1$), then $\mathcal B(x,p,l) = \mathcal B(x)$ (cf. the first case study). 
\end{remark}

\section{Computation of CBC}\label{compute_CBC}
We employ an approach based on a counter-example guided inductive synthesis (CEGIS) framework to find a CBC for each mode of subsystems
$\Sigma_{p_i}$ with a finite set of modes $P_i,~ i\in\{1,\dots,N\}$. The approach employs satisfiability (feasibility) solvers for finding CBC of a given parametric form employing existing Satisfiability Modulo Theories (SMT) solvers such as Z3 \cite{de2008z3}, dReal \cite{gao2013dreal}, and OptiMathSAT \cite{sebastiani2015optimathsat}. In order to use CEGIS framework, we raise the following assumption.

\begin{assumption}\label{ass:BC1}
	Each mode of switched subsystem
	$\Sigma_{p_i}, ~\forall p_i\in{P}, ~\forall i\in\{1,\dots,N\}$, has a compact state set $X_i$, and a compact internal input set $W_i$.
\end{assumption}

Under Assumption~\ref{ass:BC1}, conditions~\eqref{subsys1}-\eqref{csbceq} can be rephrased as
a satisfiability problem which can be searched for a parametric CBC using CEGIS approach. The feasibility condition that is required to be satisfied for the existence of a CBC $\mathcal B_{p_i}$ is given in the following lemma.

\begin{lemma}\label{sos1}
	Consider a mode of switched subsystem
	$\Sigma_{p_i}$
	satisfying Assumption~\ref{ass:BC1}. Suppose there exist functions $\mathcal B_{p_i}$, $\alpha_{p_i}\in\mathcal{K}_\infty$, $\rho_{\mathrm{int}p_i}\in\mathcal{K}_\infty\cup\{0\}$, and constants $0<\kappa_{p_i}<1$, $\gamma_{p_i},\psi_{p_i}\in\R_{\geq 0}$ and $\lambda_{p_i}\in\R_{> 0}$, such that the following expression is true:
	\begin{align}\notag
		&\bigwedge_{x_i\in X_i} \mathcal B_{p_i} \geq \alpha_{p_i}(\Vert h_i(x_i)\Vert) 
		\bigwedge_{x\in {X_0}_i} \mathcal B_{p_i} \leq \gamma_{p_i}
		\bigwedge_{x\in {X_1}_i} \mathcal B_{p_i} \geq \lambda_{p_i}\bigwedge_{x_i\in X_i}\bigwedge_{w_i\in W_i} (\mathbb{E}\Big[\mathcal B_{p_i}(x_i(k+1)) \,\,\big|\,\, x_i, w_i\Big]\\\label{eq:sos4}
		&\quad\quad\quad\quad\leq \max\Big\{\kappa_{p_i}\mathcal B_{p_i}(x_i), \rho_{\mathrm{int}p_i}(\|w_i\|),\psi_{p_i}\Big\}. 
	\end{align}
	Then, $\mathcal B_{p_i}$ satisfies conditions~\eqref{subsys1}-\eqref{csbceq} in Definition~\ref{cbc}.
\end{lemma}

Now one can utilize the CEGIS approach to search for a parametric CBC satisfying~\eqref{eq:sos4} (implying original conditions~\eqref{subsys1}-\eqref{csbceq}). For a
detailed discussion on the CEGIS approach, we refer the interested reader to~\cite[Subsection 5.3.2]{Pushpak2019}.

\section{Case Study}
To demonstrate the effectiveness of the proposed results, we first apply our approaches to a room temperature network in a circular building containing $1000$ rooms. We compositionally synthesize safety controllers to maintain the temperature of each room in a comfort zone in a bounded time horizon. Moreover, to show the applicability of our results to switched systems accepting \emph{multiple} barrier certificates with a \emph{dwell-time} condition, we apply our technique to a circular cascade network of $500$ subsystems (totally $1000$ dimensions) and provide upper bounds on the probability that the interconnected system reaches some unsafe region in a finite-time horizon.

\subsection{Room Temperature Network}
\label{example}

The model of this case study is borrowed from~\cite{Meyer.2018} by including a stochasticity in the model as an additive noise. The evolution of the temperature $T(\cdot)$ in the interconnected system is governed by the following dynamics:
\begin{equation*}
	\Sigma: \begin{cases}
		T(k+1)=AT(k) + \theta T_hB_{\bold{p}(k)} + \beta T_E + 0.25\varsigma(k), \\
		y(k)=T(k),
	\end{cases}
\end{equation*}
where $A \in \mathbb{R}^{n \times n}$ is a matrix with diagonal elements given by $\bar a_{ii}=(1-2 \eta-\beta-\theta b_{ip_i})$, off-diagonal elements $\bar a_{i,i+1}=\bar a_{i+1,i}=\bar a_{1,n}=\bar a_{n,1}= \eta$, $i\in \{1,\ldots,n-1\}$, and all other elements are identically zero. Parameters $\eta = 0.005$, $\beta = 0.022$, and $\theta = 0.05$ are conduction factors, respectively, between the rooms $i \pm 1$ and $i$, the external environment and the room $i$, and the heater and the room $i$.  Outside temperatures are the same for all rooms: $T_{ei}=-1\,{}^{\circ}\mathsf{C}$, $\forall i\in\{1,\ldots,n\}$, and the heater temperature is $T_h=50\,{}^{\circ}\mathsf{C}$. Moreover, $T(k)=[T_1(k);\ldots;T_n(k)]$, $\varsigma=[\varsigma_1(k);\ldots;\varsigma_n(k)]$, $T_E=[T_{e_1};\ldots;T_{e_n}]$, and $B_p=[b_{1p_1};\ldots;b_{np_n}]$, such that

\begin{align}\notag
	b_{ip_i} =\left\{\hspace{-1mm}\begin{array}{l} 0,\quad\quad\quad\quad\! \text{if}~~~~  p_i = 1,\\
		0.1,\quad\quad \quad\text{if}~~~~  p_i = 2,\\
		0.2,\quad\quad \quad\text{if}~~~~  p_i = 3,\\
		0.3,\quad\quad \quad\text{if}~~~~  p_i = 4,\\
		0.4,\quad\quad \quad\text{if}~~~~  p_i = 5,\\
		0.5,\quad\quad \quad\text{if}~~~~  p_i = 6,\\
		0.6,\quad\quad \quad\text{if}~~~~  p_i = 7,\\
	\end{array}\right.
\end{align}
with the finite set of modes $P_i = \{1,\dots,7\},i\in\{1,\dots,n\}$.
Now by considering the individual rooms as $\Sigma_i$ represented by
\begin{align}\notag
	\Sigma_i: \begin{cases}
		T_i(k+1) = \bar a_{ii}T_i(k) + \theta T_h b_{i{\bold{p}_i(k)}} +  \eta w_i(k) + \beta T_{ei}(k) + 0.25\varsigma_i(k), \\
		y_i(k)=T_i(k),
	\end{cases}
\end{align}
one can readily verify that $\Sigma=\mathcal{I}(\Sigma_1,\ldots,\Sigma_N)$, equivalently $\Sigma=\mathcal{I}(\mathbb{A}(\Sigma_1),\ldots,,\mathbb{A}(\Sigma_N))$, where $w_i(k) = [T_{i-1}(k);T_{i+1}(k)]$ (with $T_0 = T_n$ and $T_{n+1} = T_1$). 

The regions of interest in this example are $X_i \in [1,50], {X_0}_i \in [19,21], {X_1}_i = [1,17]\cup [23,50], \forall i\in\{1,\dots,n\}$. The main goal is to find an ABC for the interconnected system such that a switching signal is synthesized for $\Sigma$ keeping the temperature of rooms in the comfort zone $[17,23]^{1000}$.

Note that in this example $\mathcal B_p = \mathcal B_{p'}, \forall p,p' \in P$  (\emph{i.e.,} there exists a common barrier certificate with $\mu =1$). Then $\mathcal B(x,p,l) = \mathcal B(x)$ as discussed in Remark~\ref{Common_barrier}. We employ the SMT solver Z3 and CEGIS approach to compute
an APBC of an order $4$ based on Lemma~\ref{sos1} as $\mathcal B_i(T_i) = -0.00012T_i^4 + 0.01045T_i^3 - 0.19932T_i^2 -   0.64538T_i + 28.68175$. Furthermore, the corresponding constants and functions in Definition~\ref{Def_1a} satisfying conditions~\eqref{Eq_2a}-\eqref{Eq_3a} are quantified as $\gamma_i = 0.16, \lambda_i = 1.2, \psi_i = 7.07 \times 10^{-4}, \kappa_i = 0.99, \alpha_i (s) = 4.5 \times 10^{-5} s^2,$ and  $\rho_i(s) = 9.3 \times 10^{-6} s^2, \forall s\in\R_{\geq 0}$. Then $\mathcal B_i(x_i)$ is an APBC for $\mathbb{A}(\Sigma_i)$,

We now proceed with constructing an ABC for the interconnected system using APBC of subsystems.  We check the small-gain condition~\eqref{Assump: Kappa} that is required for the compositionality result. By taking $\sigma_i(s) = s$, $\forall i\in\{1,\ldots,n\}$, the condition~\eqref{Assump: Kappa} and as a result the condition \eqref{compositionality} are always satisfied. Moreover, the compositionality condition~\eqref{compositionality1} is met since $\lambda_i > \gamma_i, \forall i\in \{1,\dots,n\}$. Then one can conclude that $\mathcal  B(T) = \max_{i} \Big\{-0.00012T_i^4 + 0.01045T_i^3 - 0.19932T_i^2 -   0.64538T_i + 28.68175\Big\}$ is an ABC for $\mathbb{A}(\Sigma)$ satisfying conditions~\eqref{Eq_2}-\eqref{eq6666} with $\gamma= 0.16, \lambda = 1.2, \kappa = 0.99$, and $\psi = 7.07\times10^{-4}$.

By employing Theorem~\ref{Kushner}, one can guarantee that the temperature of the interconnected system $\Sigma$ starting from the initial condition $a \in [19,21]^{1000}$ remains in the comfort region $[17,23]^{1000}$ during the time horizon $T_d=10$ with a probability at least $87\%$, \emph{i.e.,}
\begin{equation}\label{threshold}
	\mathbb{P}\Big\{\mathcal B(T(k))< 1.2\,\,\big|\,\, a,\,\, \forall k\in[0,10]\Big\}\ge 0.87\,.
\end{equation}

\begin{figure}
	\center
	\includegraphics[width=0.55\linewidth]{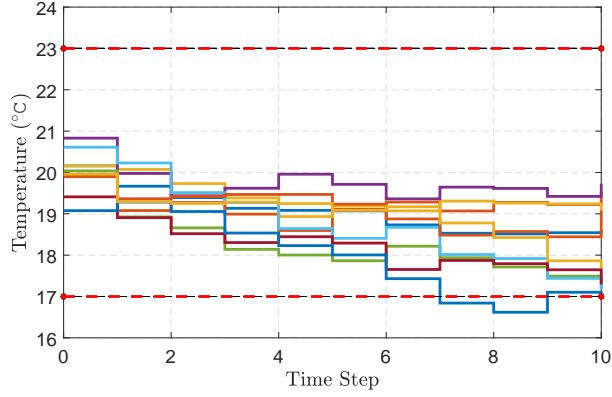}
	\caption{Closed-loop state trajectories of a representative room with $10$ noise realizations in a network of $1000$ rooms.}
	\label{trajectory}
\end{figure}

State trajectories of the closed-loop system in a network of $1000$ rooms for a representative room with $10$ noise realizations are illustrated in Figure~\ref{trajectory}.

\subsection{Switched Systems Accepting Multiple Barrier Certificates with Dwell-Time}
In order to show the applicability of our results to switched systems accepting \emph{multiple} barrier certificates with a \emph{dwell-time} condition, we apply our techniques to a circular cascade network of $500$ subsystems (totally $1000$ dimensions). The model of the system does not have a common barrier certificate because it exhibits unstable behaviors for different switching signals~\cite{liberzon2003switching} (\emph{i.e.,} if one periodically switches between different modes, the trajectory goes to infinity). The dynamics of the interconnected system are described by
\begin{equation*}
	\Sigma:\left\{\hspace{-1mm}\begin{array}{l}{x}(k+1)= A_{\bold{p}(k)}x(k)+B_{\bold{p}(k)}+R\varsigma(k),\\
		y(k)=x(k),\end{array}\right.
\end{equation*}
where
\begin{align}\notag
	A_{\bold{p}(k)}&=\begin{bmatrix}\bar A_{{pi}} & 0  & \cdots & \cdots & \tilde A  \\  \tilde A  & \bar A_{{pi}} & 0  & \cdots & 0  \\ 0  & \tilde A  & \bar A_{{pi}} & \cdots & 0  \\ \vdots &  & \ddots & \ddots & \vdots \\ 0  & \cdots & \cdots & \tilde A  & \bar A_{{pi}}\end{bmatrix}_{n\times n}\!\!\!\!\!\!\!\!\!\!\!\!,\\\notag
	\bar A_{{p_i}}& =\left\{\hspace{-1.7mm}\begin{array}{l} \begin{bmatrix}0.05 & 0 \\  0.9 & 0.03\end{bmatrix}\!,\quad~\! \quad\text{if}~~~  p_i = 1,\vspace{1.7mm}\\
		\begin{bmatrix}0.02 & -1.2 \\  0 & 0.05\end{bmatrix}\!,\quad\text{if}~~~  p_i = 2,\\
	\end{array}\right.
	\tilde A= \begin{bmatrix}0.01 & 0 \\  0 & 0.01\end{bmatrix}\!.
\end{align}
We choose $R = \mathsf{diag}(0.1\mathds{1}_{2},\ldots,0.1\mathds{1}_{2})$ and fix here $N=500$. Furthermore, $B_p=[b_{1p_1};\ldots;b_{Np_N}]$ such that
\begin{align}\notag
	b_{ip_i} =\left\{\hspace{-1.7mm}\begin{array}{l} \begin{bmatrix}-0.9 \\  0.5 \end{bmatrix}\!,\quad\quad \quad\vspace{1mm}\text{if}~~~~  p_i = 1,\\
		\begin{bmatrix}0.9 \\  -0.2 \end{bmatrix}\!,\quad\quad\quad \text{if}~~~~  p_i = 2.\\
	\end{array}\right. 
\end{align}
We partition $x(k)$ as $x(k)=[x_1(k);\ldots;x_N(k)]$ and $\varsigma(k)$ as $\varsigma(k)=[\varsigma_1(k);\ldots;\varsigma_N(k)]$, where $x_i(k),\varsigma_i(k)\in\R^{2}$, \emph{i.e.,} $x_i = [x_{i1};x_{i2}], \varsigma_i = [\varsigma_{i1};\varsigma_{i2}]$.
Now, by introducing the individual subsystems $\Sigma_i$ described as
\begin{align}\notag
	\Sigma_i:\left\{\hspace{-2mm}\begin{array}{l}x_i(k+1)= \bar A_{{\bold{p}_i(k)}}x_i(k) +\tilde A_iw_i(k)+b_{i{\bold{p}_i(k)}}+0.1\mathds{1}_{2}\varsigma_i(k),\\
		y_i(k)=x_i(k),\\
	\end{array}\right.
\end{align}
where $w_i(k)=y_{i-1}, ~i \in \{1,\ldots,N\},$ with  $y_0 = y_N$, one can readily verify that $\Sigma=\mathcal{I}(\Sigma_1,\ldots,\Sigma_N)$, equivalently $\Sigma=\mathcal{I}(\mathbb{A}(\Sigma_1),\ldots,\mathbb{A}(\Sigma_N))$.

The regions of interest here are $X_i \in [-6,6]^2, X_{0_i} \in [-0.5,0.5]^2, X_{1_i} = [-6,-2]^2\cup [2,6]^2, \forall i\in\{1,\dots,N\}$. The main goal is to find an ABC for the interconnected system such that a switching signal is synthesized for $\Sigma$ regulating the state of subsystems in a safe zone $[-2,2]^{1000}$.  We first find a CBC for each mode based on Definitions~\ref{cbc} using software tool \textsf{SOSTOOLS}~\cite{papachristodoulou2013sostools} and the SDP solver \textsf{SeDuMi}~\cite{sturm1999using}. One can verify that, $\forall i\in\{1,\ldots,N\}$, conditions~\eqref{subsys1}-\eqref{csbceq} are satisfied by
\begin{align}\notag
	\text{for} \,\,p_i = 1\!: \quad &\gamma_{{p_i}} = 0.15, \lambda_{{p_i}} = 2.4, \kappa_{{p_i}} = 0.469, \psi_{{p_i}} = 5.42\times10^{-6},  \\\notag
	& \alpha_{{p_i}}(s) = 4\times 10^{-5}s^2, \rho_{\mathrm{int}p_i}(s) = 2.71\times10^{-6}s^2, \forall s\in\R_{\geq 0},\\\notag
	\text{for} \,\,p_i = 2\!: \quad & \gamma_{{p_i}} = 0.16, \lambda_{{p_i}} = 2.3, \kappa_{{p_i}} = 0.498, \psi_{{p_i}} = 6.88\times10^{-6}, \\\notag
	&\alpha_{{p_i}}(s) = 5\times 10^{-5}s^2, \rho_{\mathrm{int}p_i}(s) = 3.44\times10^{-6}s^2, \forall s\in\R_{\geq 0},
\end{align}
with $\mathcal B_{1_i}(x_i) = 0.2309x_{i1}^2 + 0.1160 x_{i1}x_{i2}+ 0.000001x_{i1}+ 0.2529x_{i2}^2- 0.000001x_{i2}+ 0.000000002$, $\mathcal B_{2_i}(x_i) = 0.2394x_{i1}^2 + 0.1101 x_{i1}x_{i2}- 0.000002x_{i1}+ 0.2588x_{i2}^2- 0.000008x_{i2}+ 0.000000005$. One can also verify that the condition~\eqref{Barrier_rate} is met with $\mu = 2$.
By taking $\epsilon =2$, one can get the dwell-time $k_d = 3$. Hence, $\mathcal B_i(x_i,p_i,l_i)=\frac{1}{\kappa_{p_i} ^ {l/2}}\mathcal B_i(x_i)$ is an APBC for $\mathbb{A}(\Sigma_i)$ satisfying conditions~\eqref{Eq_2a}-\eqref{Eq_3a} with  $\alpha_{i}(s)=4\times 10^{-5}s^2$, $\forall s\in \mathbb R_{\ge0}, \gamma_i = 0.321, \lambda_i = 2.3, \kappa_i=0.706$, $\rho_{\mathrm{int}i}(s)=9.78\times10^{-6}s^2$, $\forall s\in \mathbb R_{\ge0}$, and $\psi_i = 1.95\times10^{-5}$.

We now proceed with constructing an ABC for the interconnected system using APBC of subsystems.  We check the small-gain condition~\eqref{Assump: Kappa}. By taking $\sigma_i(s) = s$, $\forall i\in\{1,\ldots,N\}$, the condition~\eqref{Assump: Kappa} and as a result the condition \eqref{compositionality} are always satisfied. Moreover, the compositionality condition~\eqref{compositionality1} is met since $\lambda_i > \gamma_i, \forall i\in \{1,\dots,N\}$.  Hence, $\mathcal B(x,p,l)=\max_{i} \{\frac{1}{\kappa_{p_i} ^ {l/2}}\mathcal B_i(x_i)\}$  is an ABC for the interconnected $\mathbb{A}(\Sigma)$ satisfying conditions~\eqref{Eq_2}-\eqref{eq6666} with $\gamma= 0.321, \lambda = 2.3, \kappa = 0.706$, and $\psi = 1.95\times10^{-5}$.

By employing Theorem~\ref{Kushner}, one can guarantee that the state of the interconnected system $\Sigma$ starting from the initial condition $a \in [-0.5,0.5]^{1000}$, with any initial mode $p_0$ and $l_0 = 0$,  remains in the safe set $[-2,2]^{1000}$ during the time horizon $T_d=100$ with a probability at least $86\%$, \emph{i.e.,}
\begin{equation}\notag
	\mathbb{P}\Big\{\mathcal B(x(k),p(k),l(k)) < 2.3\,\,\big|\,\, a,p_0,l_0, \forall k\in[0,100]\Big\}\ge 0.86.
\end{equation}

\section{Discussion}
In this work, we proposed a compositional scheme for constructing control
barrier certificates for large-scale stochastic switched systems accepting \emph{multiple} barrier certificates with some \emph{dwell-time} conditions. Those barrier certificates provide upper bounds on the probability that interconnected systems reach certain unsafe regions in finite-time horizons. The main goal was to synthesize control policies driving switching signals satisfying safety properties for interconnected systems by utilizing so-called augmented pseudo-barrier certificates of subsystems. We constructed augmented barrier certificates for interconnected switched systems using augmented pseudo-barrier certificates of subsystems as long as some $\max$-type small-gain conditions hold. We employed a systematic technique based on a counter-example guided inductive synthesis (CEGIS) approach and computed control barrier certificates for each mode of a subsystem. We illustrated our proposed results by applying them to two different case studies.

\section{Acknowledgment}
The authors would like to thank Abolfazl Lavaei for the fruitful discussions and helpful comments.

\bibliographystyle{alpha}
\bibliography{biblio}

\section{Appendix}
\begin{proof}\textbf{(Theorem~\ref{Kushner})}
	According to the condition~\eqref{Eq_3}, $\mathbb{X}_1\subseteq \{(x,p,l)\in \mathbb{X} \,\,\big|\,\, \mathcal B(x,p,l) \ge \lambda \}$. Then we have
	\begin{align}\notag
		\mathbb{P}&\Big\{(x(k),p(k),l(k))\in \mathbb X_1 \text{ for } 0\leq k\leq T_d\,\,\big|\,\, a,p_0,l_0\Big\}\\\label{Eq:5}
		&\leq\mathbb{P}\Big\{ \sup_{0\leq k\leq T_d} \mathcal B(x(k),p(k),l(k)) \geq \lambda \,\, \big|\,\, a,p_0,l_0 \Big\}.
	\end{align}
	The proposed bounds in~\eqref{eqlemma2} follows directly by applying~\cite[Theorem 3, Chapter III]{1967stochastic} to~\eqref{Eq:5} (but adapted
	to stochastic switched systems) and employing respectively conditions~\eqref{eq6666} and~\eqref{Eq_2}.
\end{proof}

\begin{proof}\textbf{(Theorem~\ref{Thm: Comp})}
	We first show that conditions~\eqref{Eq_2} and \eqref{Eq_3} in Definition \ref{Def_2a} hold. For any $(x,p,l) \in \mathbb{X}_0 = \prod_{i=0}^{N} \mathbb{X}_{0_i} $ and from \eqref{Eq_2a1}, we have
	\begin{align}\notag
		\mathcal B(x,p,l) = \max_{i} \Big\{\sigma_i^{-1}(\mathcal B_{i}(x_{i},p_{i},l_{i}))\Big\}\leq\max_{i} \Big\{\sigma_i^{-1}(\gamma_i)\Big\} = \gamma,
	\end{align} 
	and similarly for any $(x,p,l) \in \mathbb{X}_1 = \prod_{i=1}^{N} \mathbb{X}_{1_i} $ and from \eqref{Eq_2a2}, one has
	\begin{align}\notag
		\mathcal B(x,p,l) = \max_{i} \Big\{\sigma_i^{-1}(\mathcal B_{i}(x_{i},p_{i},l_{i}))\Big\}\geq\max_{i} \Big\{\sigma_i^{-1}(\lambda_i)\Big\} = \lambda,
	\end{align} 
	satisfying conditions \eqref{Eq_2} and \eqref{Eq_3} with $\gamma = \max_{i} \Big\{\sigma_i^{-1}(\gamma_i)\Big\}$ and $\lambda = \max_{i} \Big\{\sigma_i^{-1}(\lambda_i)\Big\}$. 
	
	Now we show that the condition~\eqref{eq6666} holds, as well. Let $\kappa(s)= \max_{i,j}\{\sigma_{i}^{-1}\circ\kappa_{ij}\circ\sigma_{j}(s)\}$. It follows from~\eqref{compositionality} that $\kappa<\mathcal{I}_d$. Moreover, $\lambda > \gamma$ according to~\eqref{compositionality1}. Since $\max_{i}\sigma_{i}^{-1}$ is concave, one can readily acquire the chain of inequalities in \eqref{Equ1b} using Jensen's inequality, and by defining the constant $\psi$ as
	\begin{align*}
		\psi:=\max_{i}\sigma_{i}^{-1}(\psi_{i}).
	\end{align*}	 
	Hence $\mathcal B(x,p,l)$ is an ABC for the interconnected augmented dt-SS $\mathcal{I}(\mathbb{A}(\Sigma_1),\ldots,\mathbb{A}(\Sigma_N))$ which completes the proof.
\end{proof}

\begin{figure*}
	\rule{\textwidth}{0.1pt}
	\begin{align}\nonumber
	\mathbb{E}&\Big[\mathcal B(x',p',l')\,\big|\,x,p,l\Big]=\mathbb{E}\Big[\max_{i}\Big\{\sigma_{i}^{-1}(\mathcal B_{i}(x'_i,p'_i,l'_i))\Big\}\,\big|\,x,p,l\Big]\\\notag
	&\le\max_{i}\Big\{\sigma_{i}^{-1}(\mathbb{E}\Big[\mathcal B_{i}(x'_i,p'_i,l'_i)\,\big|\,x,p,l\Big])\Big\}=\max_{i}\Big\{\sigma_{i}^{-1}(\mathbb{E}\Big[\mathcal B_{i}(x'_i,p'_i,l'_i)\,\big|\,x_i,p_i,l_i\Big])\Big\}\\\notag
	&\leq\max_{i}\Big\{\sigma_{i}^{-1}(\max\{\kappa_{i}\mathcal B_{i}(x_i,p_i,l_i),\rho_{\mathrm{int}i}(\Vert w_i\Vert), \psi_{i}\})\!\Big\}\!=\!\max_{i}\Big\{\sigma_{i}^{-1}(\max\{\kappa_{i}\mathcal B_{i}(x_i,p_i,l_i),\rho_{\mathrm{int}i}(\max_{j, j\neq i}\{\Vert w_{ij}\Vert\}), \psi_{i}\})\!\Big\}\\\notag
	&=\max_{i}\Big\{\sigma_{i}^{-1}(\max\{\kappa_{i}\mathcal B_{i}(x_i,p_i,l_i),\rho_{\mathrm{int}i}(\max_{j, j\neq i}\{\Vert y_{ji}\Vert\}), \psi_{i}\})\Big\}\\\notag
	&=\max_{i}\Big\{\sigma_{i}^{-1}(\max\{\kappa_{i}\mathcal B_{i}(x_i,p_i,l_i), \rho_{\mathrm{int}i}(\max_{j, j\neq i}\{\Vert \mathbb{H}_j(x_j,p_j,l_j)\Vert\}), \psi_{i}\})\Big\}\\\notag
	&\leq\max_{i}\Big\{\sigma_{i}^{-1}(\max\{\kappa_{i}\mathcal B_{i}(x_i,p_i,l_i),\rho_{\mathrm{int}i}(\max_{j , j\neq i}\{\alpha_{j}^{-1}(\mathcal B_{j}(x_j,p_j,l_j))\}),\psi_{i}\})\Big\}\\\notag
	&=\max_{i,j}\Big\{\sigma_{i}^{-1}(\max\{\kappa_{ij}\mathcal B_{j}(x_j,p_j,l_j),\psi_{i}\})\Big\}=\max_{i,j}\Big\{\sigma_{i}^{-1}(\max\{\kappa_{ij}\circ \sigma_{j}\circ \sigma_{j}^{-1}(\mathcal B_{j}(x_j,p_j,l_j)),\psi_{i}\})\Big\}\\\notag
	&\leq\max_{i,j,{z}}\Big\{\sigma_{i}^{-1}(\max\{\kappa_{ij}\circ \sigma_{j} \circ \sigma_{z}^{-1}(\mathcal B_{z}(x_{z},p_{z},l_{z})),\psi_{i}\})\Big\}\\\label{Equ1b}
	&=\max_{i,j}\Big\{\sigma_{i}^{-1}(\max\{\kappa_{ij}\circ\sigma_{j}(\mathcal B(x,p,l)),\psi_{i}\})\Big\}=\max\Big\{\kappa\mathcal B(x,p,l),\psi\Big\}.
	\end{align}
	\rule{\textwidth}{0.1pt}
\end{figure*}

\begin{proof}\textbf{(Theorem~\ref{Thm_5a})}
	For any $(x,p,l)\in \mathbb{X}$, we get 
	\begin{align}\notag
		&\Vert \mathbb{H}(x,p,l)\Vert =\Vert h(x)\Vert \leq \alpha_p^{-1}(\mathcal B_p(x))= \alpha_p^{-1}({\kappa_p}^{l/\epsilon}\, \mathcal B((x,p,l))).
	\end{align}
	Since $\frac{1}{\kappa_{p}^{l/\epsilon}}>1$, one can conclude that the inequality \eqref{Eq_2a} holds with $\alpha(s)=\min_p\{\alpha_p(s)\}$, $\forall s\in \R_{\geq0}$.
	Now we show that inequalities~\eqref{Eq_2a1} and~\eqref{Eq_2a2} hold, as well.
	For any $(x,p,l)\in \mathbb{X}_0$, one has
	\begin{align}\notag
		\mathcal B(x,p,l)=\frac{1}{{\kappa_p}^{l/\epsilon}}\mathcal B_p(x)\leq\frac{1}{{\kappa_p}^{l/\epsilon}} \gamma_p,
	\end{align}
	and similarly for any $(x,p,l)\in \mathbb{X}_1$, one has
	\begin{align}\notag
		\mathcal B(x,p,l)=\frac{1}{{\kappa_p}^{l/\epsilon}}\mathcal B_p(x)\geq\frac{1}{{\kappa_p}^{l/\epsilon}}\lambda_p ,
	\end{align}
	satisfying conditions \eqref{Eq_2a1} and \eqref{Eq_2a2} with $\gamma = \max_p\{\frac{1}{{\kappa_p}^{(k_d-1)/\epsilon}}\gamma_p\}$ and $\lambda = \min_p\{\lambda_p\}$ (since $\frac{1}{\kappa_{p}^{l/\epsilon}}>1$). 
	
	Now we proceed with showing the inequality~\eqref{Eq_3a}. In order to show that the function $\mathcal B(x,p,l)$ in~\eqref{function V} satisfies~\eqref{Eq_3a}, we should consider the  three different scenarios as in Definition~\ref{def: ASS}. For the first scenario ($l<k_d-1, p' = p$, and $l' = l+1$), we have:
	\begin{align}\notag
		\mathbb{E}& \Big[\mathcal B(x',p',l')\,\big|\,x,p,l,w\Big]
		=\frac{1}{{\kappa_{p'}}^{l'/\epsilon}}\mathbb{E} \Big[\mathcal B_{p'}(x')\,\big|\,x,p,w\Big]\\\notag
		&=\frac{1}{{\kappa_{p}}^{(l+1)/\epsilon}}\mathbb{E} \Big[\mathcal B_p(f_p(x,w,\varsigma))\,\big|\,x,w\Big]\\\notag
		&\leq\frac{1}{{\kappa_{p}}^{(l+1)/\epsilon}}\max\Big\{\kappa_p\mathcal B_p(x(k)), \rho_{\mathrm{int}p}(\|w\|),\psi_p\Big\}\\\notag
		&=\max\Big\{\kappa_p^\frac{\epsilon-1}{\epsilon}\mathcal B_p(x,p,l), \frac{1}{{\kappa}_{p}^{(l+1)/\epsilon}}\rho_{\mathrm{int}p}(\|w\|),\frac{1}{{\kappa}_{p}^{(l+1)/\epsilon}}\psi_p\Big\}\\\notag
		&\leq\max\Big\{\kappa_p^\frac{\epsilon-1}{\epsilon}\mathcal B_p(x,p,l), \frac{1}{{\kappa}_{p}^{k_d/\epsilon}}\rho_{\mathrm{int}p}(\|w\|),\frac{1}{{\kappa}_{p}^{k_d/\epsilon}}\psi_p\Big\};
	\end{align}
	Note that the last inequality holds since $l<k_d-1$, and consequently, $l+1<k_d$. 
	
	For the second scenario ($l=k_d-1, p' = p$, and $l'=k_d-1$), we have:
	\begin{align}\notag
		\mathbb{E}& \Big[\mathcal B(x',p',l')\,\big|\,x,p,l,w\Big]
		=\frac{1}{{\kappa_{p'}}^{l'/\epsilon}}\mathbb{E} \Big[\mathcal B_{p'}(x')\,\big|\,x,p,w\Big]\\\notag
		&=\frac{1}{{\kappa_{p}}^{l/\epsilon}}\mathbb{E} \Big[\mathcal B_p(f_p(x,w,\varsigma))\,\big|\,x,w\Big]\\\notag
		&\leq\frac{1}{{\kappa_{p}}^{l/\epsilon}}\max\Big\{\kappa_p\mathcal B_p(x(k)), \rho_{\mathrm{int}p}(\|w\|),\psi_p\Big\}\\\notag
		&=\max\Big\{\kappa_p\mathcal B(x,p,l), \frac{1}{{\kappa_{p}}^{l/\epsilon}}\rho_{\mathrm{int}p}(\|w\|),\frac{1}{{\kappa_{p}}^{l/\epsilon}}\psi_p\Big\}\\\notag
		&\le\max\Big\{\kappa_p^\frac{\epsilon-1}{\epsilon}\mathcal B(x,p,l), \frac{1}{{\kappa}_{p}^{k_d/\epsilon}}\rho_{\mathrm{int}p}(\|w\|),\frac{1}{{\kappa}_{p}^{k_d/\epsilon}}\psi_p\Big\};\notag
	\end{align}
	Note that the last inequality holds since $\epsilon > 1$, and consequently, $0<\frac{\epsilon-1}{\epsilon}<1$.
	
	For the last scenario ($l=k_d-1, p'\neq p$, and $l' = 0$), using Assumption~\ref{Assume: Switching} we have:
	\begin{align}\notag
		\mathbb{E}& \Big[\mathcal B((x',p',l'))\,\big|\,x,p,l,w\Big]
		=\frac{1}{{\kappa_{p'}}^{l'/\epsilon}}\mathbb{E} \Big[\mathcal B_{p'}(x')\,\big|\,x,p,w\Big]\\\notag
		&\leq\mu\,\mathbb{E} \Big[\mathcal B_p(f_p(x,w,\varsigma))\,\big|\,x, w\Big]\\\notag
		&\leq\mu\max\Big\{\kappa_p\mathcal B_p(x(k)), \rho_{\mathrm{int}p}(\|w\|),\psi_p\Big\}\\\notag
		&=\mu{\kappa_{p}}^{(k_d-1)/\epsilon}\frac{1}{{\kappa_{p}}^{l/\epsilon}}\max\Big\{\kappa_p\mathcal B_p(x(k)), \rho_{\mathrm{int}p}(\|w\|),\psi_p\Big\}\\\notag
		&=\max\Big\{\mu\kappa_{p}^{(k_d-1)/\epsilon} \kappa_p\mathcal B((x,p,l)), \mu\rho_{\mathrm{int}p}(\|w\|),\mu\psi_p\Big\}\\\notag
		&\leq\max\Big\{\kappa_p\mathcal B((x,p,l)), \mu\rho_{\mathrm{int}p}(\|w\|),\mu\psi_p\Big\}\\\notag
		&\le\max\Big\{\kappa_p^\frac{\epsilon-1}{\epsilon}\mathcal B((x,p,l)), \frac{1}{{\kappa}_{p}^{k_d/\epsilon}}\rho_{\mathrm{int}p}(\|w\|),\frac{1}{{\kappa}_{p}^{k_d/\epsilon}}\psi_p\Big\};
	\end{align}
	
	Note that the last scenario holds since $\forall p\in P$, $k_d \geq \epsilon\frac{\ln(\mu)}{\ln(1/\kappa_{p})}+1$, and equivalently $\forall p\in P$, $\mu\kappa_{p}^{(k_d-1)/\epsilon}\leq 1$. By defining $\kappa =\max_p\{ \kappa_p^\frac{\epsilon-1}{\epsilon}\}$, $\rho_{\mathrm{int}}(s) =\max_p\{\frac{1}{{\kappa}_{p}^{k_d/\epsilon}}\rho_{\mathrm{int}p}(s)\},\forall s\in\mathbb R_{\geq0},$ and $\psi = \max_p\{\frac{1}{{\kappa}_{p}^{k_d/\epsilon}} \psi_p\}$, the inequality \eqref{Eq_3a} holds. Hence, $\mathcal B(x,p,l)$ is an APBC for $\mathbb{A}(\Sigma)$, which completes the proof.  		
\end{proof}

\end{document}